\def\brr#1{\left(#1\right)}
\def\brf#1{\left\{#1\right\}}
\def\z{{\bf z}}
\def\u{{\bf u}}
\def\sign{\mathop{\rm sign}\nolimits}
\def\bbar#1{\overline{#1}}

\documentclass[12pt]{article}

\usepackage{amsfonts}
\usepackage{amsmath}
\usepackage{amsthm}
\usepackage[dvips]{color}
\usepackage[dvips]{graphicx}
\newtheorem{prop}{Proposition}

\begin{document}

\begin{titlepage}
\title{Star graphs and \v Seba billiards}
\author{G.~Berkolaiko$^1$, E.B.~Bogomolny,$^2$ and
  J.P.~Keating$^{3,4}$\\
  \\
  \small\it
  $^1$Dept. of Physics of Complex Systems, Weizmann Institute of
  Science,\\ 
  \small\it Rehovot 76100, Israel\\
  \small\it
  $^2$Laboratoire de Physique Th\'eorique et Mod\`eles Statistiques, 
  Universit\'e Paris-Sud,\\ 
  \small\it 91405 Orsay Cedex, France\\
  \small\it
  $^3$ School of Mathematics, University of Bristol, Bristol BS8 1TW,
  UK\\
  \small\it
  $^4$BRIMS, Hewlett-Packard Laboratories, Filton Road, Stoke
  Gifford,\\ 
  \small\it Bristol BS34 8QZ, UK}
\end{titlepage}

\maketitle
\begin{abstract}
We derive an exact expression for the two-point correlation function
for quantum star graphs in the limit as the number of bonds tends to
infinity.  This turns out to be identical to the corresponding result
for certain \v Seba billiards in the semiclassical limit.  Reasons for this are
discussed.  The formula we derive is also shown to be equivalent to a
series expansion for the form factor --- the Fourier transform of the
two-point correlation function --- previously calculated using periodic
orbit theory.
\end{abstract}
\thispagestyle{empty}
\newpage

\section{Introduction}
The statistical distribution of quantum energy levels is a much
studied topic.  It has been conjectured that generic, classically
integrable systems give rise to uncorrelated quantum spectra
\cite{BT}, while the energy levels of generic classically chaotic
systems have the same statistical properties as the eigenvalues of
random matrices \cite{BohGS}.  This has been confirmed by
semiclassical theory \cite{Berry,Bog-Kea}, and in a large number of 
numerical studies, but classes of systems have also been found for
which it is not true; these include geodesic motion on surfaces of
constant negative curvature \cite{BGGS}, and the cat maps \cite{Kea}.

Quantum graphs \cite{KS1,KS2} are mathematical models introduced in order 
to explore the connection between the periodic orbits of a system and
the statistical properties of its energy levels.  The trace formula,
in which the level density is connected to
a sum over periodic orbits, is exact for graphs, rather than a semiclassical
approximation, and the orbits can be classified straightforwardly.  
However, despite the fact that numerical computations have
revealed good conformance of the spectral statistics of many quantum graphs to
the predictions of Random Matrix Theory (RMT), few conclusive
analytical results have been obtained so far.  This is due to
the fact that although some individual finite graphs can be shown to
reproduce certain features of RMT behaviour \cite{SS,Tan,Gas}, the
full RMT results can only be recovered in a limit in which one is
forced to consider larger and larger graphs, and this necessitates
finding general, combinatorial asymptotic techniques for dealing with
the (non-trivial) length degeneracies of the periodic orbits.

\begin{figure}
  \label{fig:star}
  \vskip 7cm
  \includegraphics{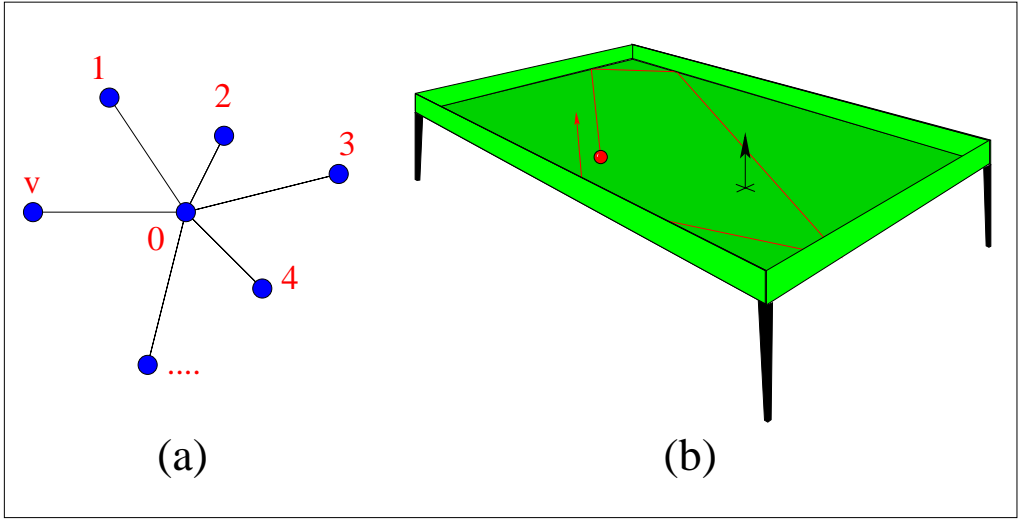}
  \caption{A star graph with $v$ edges (a) and a \v Seba billiard (b).}
\end{figure}

One family of graphs in which this goal has been achieved are the star 
graphs \cite{BK} (defined below and shown in Fig.~\ref{fig:star}), 
but in this case the resulting spectral statistics are neither RMT nor
Poissonian (i.e.~those of random numbers). 
It turns out, however, that it is not the first time that such
statistics have arisen in the connection with the study of quantum
chaos.  
Our purpose here is to demonstrate that the star graphs have
exactly the same two point spectral correlations as a large class of
quantum systems, which we will refer to as {\em \v Seba billiards}.
 
The original \v Seba billiard, a rectangular quantum billiard
perturbed by a point singularity (also illustrated in
Fig.~\ref{fig:star}), was introduced in \cite{S} as an example of a 
system whose classical counterpart is integrable (the singularity
affects only a set of measure zero of the orbits) but which nonetheless
exhibits properties of quantum chaos.  This construction was later
generalized to all integrable systems \cite{AS} perturbed in the same
way.  We will refer to any
system in this class as a \v Seba billiard.

The energy levels of a \v Seba billiard can be found by solving an
explicit equation which depends on the levels of the
original unperturbed system and on the boundary conditions imposed 
at the singularity.  This equation takes the general form 
\begin{equation}
\label{eq:Sebacond}
\lambda\xi(z)=1,
\end{equation} 
where $\xi(z)$ is the meromorphic function
\begin{equation}
  \label{eq:SE_Seba}
  \xi(z) = \sum_n \frac{|\psi_n({\bf x_0})|^2}{E_n-z},
\end{equation}
the sum being suitably regularized to ensure convergence.  Here
$\brf{E_i}$ are the eigenvalues of the unperturbed system, $\psi_n({\bf x_0})$
is the value of the $n$th unperturbed eigenfunction at the position 
${\bf x_0}$ of the singularity, and the coupling constant
$\lambda$ parametrizes the boundary conditions \cite{S,AS}.
Assuming that $\brf{E_i}$ are given
by a Poisson process, one can then calculate the associated 
spectral statistics, such as
the joint level distribution,
asymptotics of the level spacing distribution \cite{AS}, and the
two-point spectral correlation function \cite{BGS}.  The results
show the presence of spectral correlations but are
substantially different from the RMT forms.

Here we apply the methods developed for \v Seba billiards in \cite{BGS}
to calculate the two-point  
spectral correlation function for star graphs, starting from an expression
which is analogous to (\ref{eq:SE_Seba}).  The formula obtained will
be shown to be a resummation of the expansion computed from the
periodic orbit sum in \cite{BK}.  Our main result will be that 
this correlation function is the same as that already found for \v Seba
billiards in the case when $|\psi_n({\bf x_0})|^2={\rm constant}$ (e.g.~when 
the billiard is rectangular with periodic boundary conditions) and 
$\lambda \rightarrow \infty$.  We finish with a discussion of reasons 
for this coincidence.

%%%%%%%%%%%%%%%%%%%%%%%% New Section %%%%%%%%%%%%%%%%%%%%%%%%%%%%

\section{Quantum star graphs}
Star graphs are metric graphs of the
type shown on Fig.~\ref{fig:star} with a Schr\"odinger equation 
\begin{equation}
  \label{eq:schrod}
  -\frac{d^2}{d x_j^2} \Psi_j = k^2 \Psi_j, \quad x_j\in[0,L_j],
\end{equation}
defined on the bonds and boundary conditions, for example 
\begin{gather}
  \Psi_j(0) = \Psi_k(0), \label{eq:BC1}\\
  \sum_j \frac{\partial}{\partial x_j} \Psi_j(0) = 0, \label{eq:BC2}\\
  \frac{\partial}{\partial x_j} \Psi_j(L_j) = 0,
  \label{eq:BC3}
\end{gather}
specified on the vertices.  Here $L_j$ is the length of
the $j$-th bond, $j=1\ldots v$, and the real variable $x_j$
varies from 0 to $L_j$,
with 0 corresponding to the central vertex and $L_j$  
to the outer vertex.  The lengths $L_j$ are assumed to be
incommensurate; see \cite{BK} for further details.  
We refer to positive values of the parameter $k$
for which the system~(\ref{eq:schrod})-(\ref{eq:BC3}) is solvable as
eigenvalues of the quantum star graph.

Denoting the ordered sequence of eigenvalues by
$\brf{k_i}_{i=1}^\infty$, we define the spectral density by
\begin{equation}
  \label{eq:def_dens}
  d(k)=\sum_{i=1}^\infty \delta(k-k_i).  
\end{equation}
The statistic we shall mainly be concerned with is the two-point 
correlation function
\begin{equation}
  \label{eq:def_r2}
  R_2(x) = \frac1{\bbar{d}^2}\left\langle
  d(k)d\brr{k+\frac{x}{\bbar{d}}} \right\rangle-\delta(x), 
\end{equation}
where $\bbar{d} = \langle d(k) \rangle$ is the mean density, $\delta(x)$ is
the Dirac $\delta$-function, and the
average $\langle\ \cdot\ \rangle$ is either over $k$, or over the bond
lengths $L_j$ (we shall specify which in each particular context).
$R_2(x)$ is an even function and hence so is its Fourier transform,
\begin{equation}
  \label{eq:def_Ktau}
  K(\tau) = 1+2\Re \int_0^\infty (R_2(x)-1)e^{2\pi ix\tau} d\tau,
\end{equation}
which is called the form factor.

\begin{figure}[t]
\vskip 8.5cm
\includegraphics{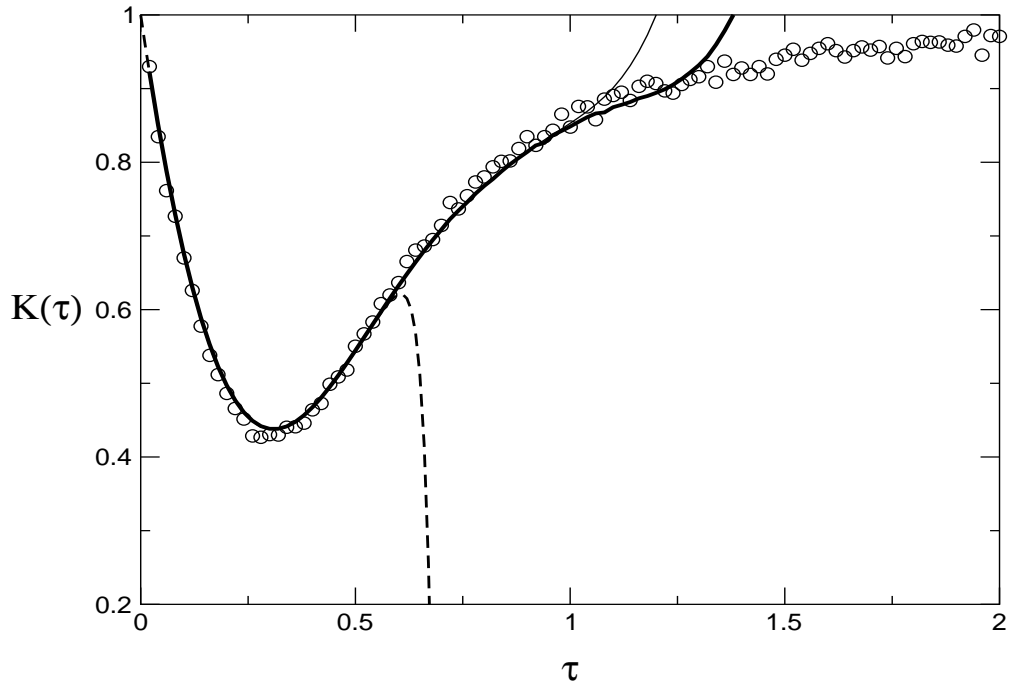}
\caption{The  sum of the first 30 terms in the expansion~(\ref{Ktau})
  (dashed line), which converges in the range
  $\tau\le\tau^*\approx0.63$, compared to the results of a numerical
  computation \cite{KS2} of $K(\tau)$ (circles).  Also shown
  are the Pad\'e approximations to the series of order 
  $[21/20]$ (thin
  solid line) and $[23/23]$ (thick solid line).} 
\label{fig:pade}
\end{figure}

A complete series expansion of the $v \rightarrow \infty$ limit of
$K(\tau)$ in powers of $\tau$ around $\tau=0$ was derived  
for the star graphs in \cite{BK} using the trace formula and a
classification of the periodic orbits:
\begin{equation}
 \label{Ktau}
  K(\tau) = \exp(-4\tau) + \sum_{j=2}^\infty\sum_{M=0}^\infty
  \frac{4^j}{j!}C_{j,M} \tau^{M+j+1},
\end{equation}
where
\begin{equation}
  C_{j,M} = (-2)^M \sum_{K=0}^M  \frac{(K+j-1)!(M-K+j-1)!}{(M+j-1)!}
  F_j(K,M-K), 
 \label{eq:CM1}
\end{equation}
with
\begin{equation}
  \label{eq:notation_F}
  F_1(K,N) = \frac{\binom{K+N}{N}}{(N+1)!(K+1)!},
\end{equation}
and
\begin{equation}
  \label{eq:FKN}
  F_j(K,N) = \sum_{k=0}^K\sum_{n=0}^N F_1(k,n)F_{j-1}(K-k,N-n).
\end{equation}
Explicitly,
\begin{equation}
  \label{eq:Ktau_firstfew}
  K(\tau) = e^{-4\tau} + 8\tau^3 - \frac{32}{3}\tau^4 +
  \frac{16}{3}\tau^5 - \frac{128}{15}\tau^6 + \frac{16}{9}\tau^7 +
  \frac{64}{63}\tau^8 + o(\tau^8).
\end{equation}
In this calculation, the average in (\ref{eq:def_r2}) was over $k$.
The result is in excellent agreement with the
numerical data (see Fig.~\ref{fig:pade}) but is limited by the fact that 
the radius of
convergence of the series is finite, being approximately $0.63$ (found
by applying Cauchy's test to the coefficients in the series, but see also
Fig.~\ref{fig:pade}).  The 
range of convergence can be extended using Pad\'e
approximation (again, see Fig.~\ref{fig:pade}), which suggests that the
singularity causing the divergence is not on the positive real line
\cite{thesis}.

Here we approach the problem from a different direction:
it is possible to solve equations (\ref{eq:schrod})-(\ref{eq:BC3}) to
derive an explicit condition on $k$ to be an eigenvalue.
Indeed, the general solution of (\ref{eq:schrod}) on a star graph
can be written in the form $\Psi_j(x) = A_j\cos(k(x+\phi_j))$, 
$j=1,\ldots,v$.  Applying condition (\ref{eq:BC3}), we obtain
$\phi_j=-L_j$ while condition~(\ref{eq:BC1}) on the central vertex
implies $A_j\cos(L_jk) = \mbox{const}$.  Finally, applying 
condition~(\ref{eq:BC2}) and dividing by $A_j\cos(L_jk)$ we obtain
\begin{equation}
  \label{eq:SE}
  \sum_{j=1}^v \tan{L_jk} = 0.
\end{equation}
Similar expressions can easily be found when different boundary
conditions are applied at the central vertex.  The general equation
reads 
\begin{equation}
  \label{eq:SE_gen}
  \sum_{j=1}^v \tan{L_jk} = \frac{1}{\lambda},
\end{equation}
where $\lambda$ is arbitrary parameter.  However, in the limit as
$v\to \infty$, $\lambda$ fixed, the two-point correlation function turns 
out to be independent of $\lambda$ (see the comment following equation 
(\ref{eq:ren_r2})).  Our calculations
will therefore be performed for $\lambda^{-1}=0$.  

Note the similarity between (\ref{eq:SE_gen}) and the quantization
condition (\ref{eq:Sebacond}) for \v Seba billiards when 
$|\psi_n({\bf x_0})|^2={\rm constant}$.

Condition (\ref{eq:SE}) means that $k$ is an eigenvalue if and only if
it is a zero of the function $F(k) = \sum_{j=1}^v \tan L_jk$, and so
we can express the density $d(k)$ as 
\begin{equation}
  \label{eq:dens_int}
  d(k) = \frac1{2\pi}\int |F'(k)| e^{izF(k)} dz = \frac1{2\pi}\int
    \sum_{s=1}^v\frac{L_s}{\cos^2{L_sk}} e^{iz\sum_{j=1}^v \tan{L_jk}}
    dz.  
\end{equation}
Our analysis of the spectral correlations will be based on this representation.

%%%%%%%%%%%%%%%%%%%%%%%%% New Section %%%%%%%%%%%%%%%%%%%%%%%%%%%%%%
\section{Mean density.}

As an example of the techniques to be employed later, we begin by 
calculating the mean density $\bbar{d}$
defined as
\begin{equation}
  \label{eq:dens_def}
  \bbar{d} = \lim_{\Delta L \to 0, {k\to \infty}} \Big\langle d(k)
  \Big\rangle_{\brf{L_j}}
\end{equation}
where now the average is with
respect to the individual lengths of the bonds, rather than over $k$:
\begin{equation}
  \label{eq:aver}
  \langle\  \cdot\  \rangle_{\brf{L_j}} = \int_{L_0}^{L_0+\Delta L}
  \hspace{-5mm}\cdots \int_{L_0}^{L_0+\Delta L}\cdot\hspace{5mm}
  \frac{dL_1}{\Delta L}\cdots\frac{dL_v}{\Delta L}.
\end{equation}
That is, we assume that the lengths are independent random variables
distributed uniformly on the interval $[L_0, L_0+\Delta L]$.  We also
assume that $\Delta L$ and $k$ tend to their respective limits in such
a way that $\Delta L k\to\infty$.

Applying this averaging to (\ref{eq:dens_int}) we obtain
\begin{eqnarray}
  \Big\langle d(k) \Big\rangle_{\brf{L_j}}\hspace{-3mm} 
  &=& \frac1{2\pi} \int_{-\infty}^\infty dz \sum_{s=1}^v \int
  \cdots \int_{L_0}^{L_0+\Delta L} L_s\frac{e^{iz\sum_{j=1}^v
      \tan{kL_j}}}{\cos^2{kL_s}} \frac{dL_1}{\Delta L} \cdots
  \frac{dL_v}{\Delta L} \nonumber\\
  &=& \frac{v}{2\pi} \int_{-\infty}^\infty dz
  \brr{\int_{L_0}^{L_0+\Delta L}\hspace{-0.3cm}e^{iz\tan{kL}}
    \frac{dL}{\Delta L}}^{v-1} 
  \brr{\int_{L_0}^{L_0+\Delta L} \hspace{-0.3cm}
    L\frac{e^{iz\tan{kL}}}{\cos^2{kL}} \frac{dL}{\Delta L}}  \nonumber\\
  \label{eq:calc_av}
  &\equiv& \frac{v}{2\pi} \int_{-\infty}^\infty {\tilde f}^{v-1}(z) 
  {\tilde g}(z) \ dz.
\end{eqnarray}
Here 
\begin{equation}
  \label{eq:g_av}
  {\tilde g}(z) = \int_{L_0}^{L_0+\Delta L}L\frac{e^{iz\tan{kL}}}{\cos^2{kL}}
    \frac{dL}{\Delta L}  \approx \frac{L_0}{\Delta L k}
    \int_{\tan{kL_0}}^{\tan{k(L_0+\Delta L)}} e^{iz\tan{kL}}\ d\tan{kL},
\end{equation}
where we were able to approximate $L$ by $L_0$ because it is slowly
varying (compared with $\tan{kL}$) and ultimately we will take the 
limit $\Delta L \to 0$.  Now, since
$\tan{kL}$ is a periodic function with the period of $\pi/k$, and the
integration is performed over the interval containing approximately
$\Delta L k/\pi$ periods, we can further approximate
\begin{eqnarray}
  \label{eq:g_av2}
  {\tilde g}(z) &=& \frac{L_0}{\Delta L k}
  \brr{\frac{\Delta L k}{\pi}\int_{-\infty}^{\infty}
    e^{iz\tan{kL}}\ d\tan{kL} + O(1)} \nonumber\\
  &\approx& 2L_0\delta(z),
\end{eqnarray}
where $O(1)$ is a quantity which is bounded as $k\Delta L \to\infty$.
Similarly, 
\begin{eqnarray}
  \label{eq:f_av}
  {\tilde f}(z) &=& \int_{L_0}^{L_0+\Delta L} e^{iz\tan{kL}} \frac{dL}{\Delta
    L} = \frac{L_0}{\Delta L k} \int_{\tan{kL_0}}^{\tan{k(L_0+\Delta L)}}
  e^{iz\tan{kL}} \frac{d\tan{kL}}{1+\tan^2{kL}} \nonumber\\
  &\approx& \frac1{\pi} \int_{-\infty}^\infty
  \frac{e^{iz\alpha}}{1+\alpha^2} d\alpha = e^{-|z|}, 
\end{eqnarray}
where the last integral was evaluated by closing the contour in either the
upper ($z>0$) or lower ($z<0$) half-plane.

Substituting the results into (\ref{eq:calc_av}) we obtain for the
average density
\begin{equation}
  \label{eq:av_d_res}
  \bbar{d} = \frac{v}{2\pi}2L_0\int_{-\infty}^\infty
  e^{-(v-1)|z|}\delta(z)dz = \frac{L_0v}{\pi},
\end{equation}
which coincides with the result of averaging over $k$ with the
bond-lengths fixed \cite{KS1,KS2,BK}.

%%%%%%%%%%%%%%%%%%%%%%%%% New Section %%%%%%%%%%%%%%%%%%%%%%%%%%%%%
\section{Two-point correlation function}
The two-point correlation function is
given by
\begin{equation}
  \label{eq:twopntnorm}
  R_2(x) = \lim_{\Delta L \to 0, k\to\infty} \frac1{{\bbar{d}}^2}
  R\brr{k,k+\frac{x}{\bbar{d}}}, 
\end{equation}
where $\bbar{d}$ is the mean density, the limit is taken in such a way
that $k\Delta L \to\infty$, and we take
\begin{eqnarray}
  \label{eq:twopnt}
  R(k_1,k_2) &=& \langle d(k_1) d(k_2) \rangle_{\brf{L_j}} \\
  \nonumber
  &=& \left\langle  \int_{-\infty}^\infty \sum_{r,s=1}^v
    \frac{L_rL_s e^{i\sum_{j=1}^v(z_1\tan{k_1L_j}+z_2\tan{k_2L_j})}}
    {\cos^2k_1L_r\cos^2k_2L_s}  
     \frac{d\z}{4\pi^2} \right\rangle_{\brf{L_j}},
\end{eqnarray}
with $\z = (z_1,z_2)$.

In this case, the analogue of (\ref{eq:calc_av}) is that
\begin{eqnarray}
  \label{eq:twopntred}
  R(k_1,k_2) = \int_{-\infty}^\infty 
  \left\{vg(\z)f^{v-1}(\z) 
  + v(v-1)\phi_1(\z)\phi_2(\z)f^{v-2}(\z)\right\} \frac{d\z}{4\pi^2},
\end{eqnarray}
where
\begin{eqnarray}
 \label{eq:intf}
  f(\z) &=& \frac1{\Delta L}\int_{L_0}^{L_0+\Delta L}
  e^{i(z_1\tan(k_1L)+z_2\tan(k_2L))} dL,\\ 
 \label{eq:intg}
  g(\z) &=& \frac1{\Delta L}\int_{L_0}^{L_0+\Delta L}
  \frac{L^2}{\cos^2k_1L\cos^2k_2L}
  e^{i(z_1\tan(k_1L)+z_2\tan(k_2L))} dL,\\ 
 \label{eq:intphi1}
  \phi_1(\z) &=& \frac1{\Delta L}\int_{L_0}^{L_0+\Delta L}
  \frac{L}{\cos^2k_1L} e^{i(z_1\tan(k_1L)+z_2\tan(k_2L))} dL,\\
 \label{eq:intphi2}
  \phi_2(\z) &=& \frac1{\Delta L}\int_{L_0}^{L_0+\Delta L}
  \frac{L}{\cos^2k_2L} e^{i(z_1\tan(k_1L)+z_2\tan(k_2L))} dL.
\end{eqnarray}

Substituting $k_1=k$, $k_2=k+\pi x/(vL_0)$, where $x$ 
is fixed, and taking
the limits 
$k\to \infty$, $\Delta L\to 0$ (while $k\Delta L \to \infty$),  we
obtain for the first integral 
\begin{eqnarray}
  f(\z) &=& \frac1{\Delta L}\int_{L_0}^{L_0+\Delta L}
  e^{i\brr{z_1\tan(kL)+z_2\tan\brr{kL + \frac{\pi xL}{vL_0}}}} dL
  \nonumber\\
  \label{eq:f_step1}
  &\approx&
  \frac1\pi \int_{-\pi/2}^{\pi/2} e^{i\brr{z_1\tan\phi+z_2\tan\brr{\phi +
      \frac{\pi x}{v}}}} d\phi,
\end{eqnarray}
where we have again used  $L/L_0\approx 1$ and, as in the
transition from (\ref{eq:g_av}) to (\ref{eq:g_av2}), we have
approximated $f$ by the integral over one period.  We now write
\begin{equation}
  \label{eq:tans}
  \tan\brr{\phi+\frac{\pi x}{v}} = \frac{\tan\phi + \tan\brr{\frac{\pi
        x}{v}}} {1-\tan\phi\tan\brr{\frac{\pi x}{v}}} = -\beta +
  \frac{1+\beta^2}{\beta-\tan\phi},
\end{equation}
where $\beta = (\tan(\pi x/v))^{-1} \propto v/(\pi x)$ (we are
interested in the $v\to\infty$ limit).  Performing the change of variables
$\alpha=\tan\phi-\beta$, we arrive at 
\begin{equation}
  \label{eq:bessel}
  f(\z) \approx \frac{e^{i\beta(z_1-z_2)}}{\pi}
  \int_{-\infty}^\infty e^{iz_1\alpha - iz_2\frac{\beta^2+1}{\alpha}}
  \frac{d\alpha}{(\alpha+\beta)^2+1}.
\end{equation}
Note that $f(\z)$ is invariant under the exchange $z_1\leftrightarrow
z_2$ and $\beta\to-\beta$, which can be verified by the change of
variables $\alpha=(\beta^2+1)/y$ in (\ref{eq:bessel}).

To evaluate the integral in (\ref{eq:bessel}) we
differentiate it with respect to $z_1$ and $z_2$ to get
\begin{eqnarray}
  \frac{\partial f}{\partial z_1} - \frac{\partial f}{\partial z_2}
  \!&=&\!    \frac{ie^{i\beta(z_1-z_2)}}{\pi}
  \int_{-\infty}^\infty e^{iz_1\alpha - iz_2\frac{\beta^2+1}{\alpha}}
  \brr{2\beta+\alpha+\frac{\beta^2+1}{\alpha}}\frac{d\alpha}{(\alpha+\beta)^2+1}\nonumber\\
  \label{eq:diff}
  &=&\! \frac{ie^{i\beta(z_1-z_2)}}{\pi}
  \int_{-\infty}^\infty e^{iz_1\alpha - iz_2\frac{\beta^2+1}{\alpha}}
  \frac{d\alpha}{\alpha} = -e^{i\beta(z_1-z_2)} \Phi(z_1,z_2),
\end{eqnarray}
where
\begin{eqnarray}
  \label{eq:Phi_def}
  \Phi(z_1,z_2) &\equiv& -\frac{i}{\pi}\int_{-\infty}^\infty
  e^{iz_1\alpha - iz_2\frac{\beta^2+1}{\alpha}} \frac{d\alpha}{\alpha}\\
  &=& 2\sign(z_1)H(-z_1z_2)
  J_0\brr{2\sqrt{-(\beta^2+1)z_1z_2}}, \nonumber
\end{eqnarray}
$J_0(x)$ being the Bessel function of the first kind and $H(x)$
the Heaviside function (characteristic function of the half axis
$[0,\infty)$).  

Applying the method of characteristics to the PDE
\begin{equation}
  \label{eq:pde}
  \frac{\partial f}{\partial z_1} - \frac{\partial f}{\partial z_2} =
  -e^{i\beta(z_1-z_2)} \Phi(z_1,z_2), 
\end{equation}
we obtain the solution
\begin{equation}
  \label{eq:ans_f}
  f(\z) = e^{-|z_1+z_2|} - \int_0^{z_1}
  e^{i\beta(2y-z_1-z_2)}\Phi\brr{y,z_1+z_2-y}dy.
\end{equation}

\medskip
Treating the integral for $g(\z)$ (see (\ref{eq:intg})) in a
fashion similar to the one used to obtain (\ref{eq:bessel}) leads us to
\begin{eqnarray}
  \label{eq:bessel_g}
  g(z_1,z_2) 
  &\approx& \frac{L_0^2}{\pi}\int_{-\pi/2}^{\pi/2}
  \frac{e^{i(z_1\tan(\phi)+z_2\tan(\phi+\pi x/v))}}
  {\cos^2(\phi)\cos^2(\phi+\pi x/v)} d\phi \\
  &=&
  L_0^2\frac{e^{i\beta(z_1-z_2)}}{\pi}
  \int_{-\infty}^\infty e^{iz_1\alpha - iz_2\frac{\beta^2+1}{\alpha}}
  \brr{1+\brr{\frac{1+\beta^2}{\alpha}+\beta}^2} d\alpha.\nonumber
\end{eqnarray}
Comparing this integral to the one in
(\ref{eq:Phi_def}), and noting that 
\begin{equation}
  \label{eq:rubbish}
  1+\brr{\frac{1+\beta^2}{\alpha}+\beta}^2 =
  \frac{\beta^2+1}{\alpha}\brr{\alpha+\beta+\frac{\beta^2+1}{\alpha}+\beta},
\end{equation}
we have that
\begin{equation}
  \label{eq:rep_g}
  g(\z) = L_0^2(\beta^2+1)\brr{\frac{\partial}{\partial z_1} -
    \frac{\partial}{\partial z_2}}\left[ e^{i\beta(z_1-z_2)}
    \Phi(z_1,z_2) \right].
\end{equation}

One can derive a similar expression for the functions $\phi_1(\z)$,
\begin{equation}
  \label{eq:rep_phi1}
  \phi_1(\z) \approx L_0 \frac{e^{i\beta(z_1-z_2)}}{\pi}
  \int_{-\infty}^\infty e^{iz_1\alpha - iz_2\frac{\beta^2+1}{\alpha}} d\alpha = L_0
  e^{i\beta(z_1-z_2)} \frac{\partial}{\partial z_1} \Phi(z_1,z_2),
\end{equation}
and $\phi_2(\z)$,
\begin{eqnarray}
  \label{eq:rep_phi2}
  \phi_2(\z) &\approx& L_0 \frac{e^{i\beta(z_1-z_2)}}{\pi}
  \int_{-\infty}^\infty e^{iz_1\alpha - iz_2\frac{\beta^2+1}{\alpha}}
  \frac{(\beta^2+1)d\alpha}{\alpha^2}\nonumber\\ 
  &=& 
  - L_0 e^{i\beta(z_1-z_2)} \frac{\partial}{\partial z_2} \Phi(z_1,z_2) .
\end{eqnarray}

Now we have all the ingredients necessary for evaluating the
integral in (\ref{eq:twopntred}). 
Substituting the expression for $g(\z)$, (\ref{eq:rep_g}),
into the first half of the integral and integrating it by parts we obtain
\begin{eqnarray}
  \label{eq:int_bp}
  \int\frac{d\z}{4\pi^2} v f^{v-1}g 
  &=&vL_0^2\int \frac{d\z}{4\pi^2} f^{v-1}(\beta^2+1)
  \brr{\frac{\partial}{\partial z_1} 
    - \frac{\partial}{\partial z_2}}\left[ e^{i\beta(z_1-z_2)}
    \Phi \right]\nonumber\\
  &=& -v L_0^2\int \frac{d\z}{4\pi^2} (\beta^2+1) e^{i\beta(z_1-z_2)}
    \Phi \brr{\frac{\partial}{\partial z_1} -
    \frac{\partial}{\partial z_2}}\left[ f^{v-1}(\z)\right] \nonumber\\
  &=& v(v-1) L_0^2\int \frac{d\z}{4\pi^2} (\beta^2+1) f^{v-2}
    e^{2i\beta(z_1-z_2)} \Phi^2.
\end{eqnarray}
Thus
\begin{equation}
  \label{eq:r2_good}
  R_2(x) = \frac{v(v-1) L_0^2}{\bbar{d}^2}\int \frac{d\z}{4\pi^2}  f^{v-2}
  e^{2i\beta(z_1-z_2)} \left[ (\beta^2+1) \Phi^2 -
    \frac{\partial \Phi}{\partial z_1} \frac{\partial \Phi}{\partial
  z_2}\right]. 
\end{equation}
Now we need to take the limit $v\to\infty$.  To do so
we write $f^{v-2}(\z) = e^{(v-2)\ln f}$ and rescale $f(\z)$
\begin{equation}
  \label{eq:f_rescaled}
  f(\u/\beta) = e^{-\frac{|u_1+u_2|}{\beta}} - \frac1{\beta}
  \int_0^{u_1} e^{i(2y-u_1-u_2)} \Psi(y,u_1+u_2-y)dy,
\end{equation}
and hence, to the leading order in $1/\beta = \pi x/v$, we have
\begin{eqnarray}
  \label{eq:lnf}
  (v-2)\ln f(\u) &\approx&
  -\pi x\brr{|u_1+u_2|+\int_0^{u_1}e^{i(2y-u_1-u_2)}
    \Psi(y,u_1+u_2-y)dy}\nonumber\\ 
  &\equiv& -\pi xQ,
\end{eqnarray}
where $\Psi$ is the rescaled function $\Phi$,
\begin{equation}
  \label{eq:Psi}
  \Psi(\u) = \Phi\brr{\frac{\u}\beta} =
  2\sign(u_1)H(-u_1u_2) J_0\brr{2\sqrt{-u_1u_2}},
\end{equation}
and we have taken the limit $v\to\infty$ ($\beta\to\infty$).

Renormalizing the rest of (\ref{eq:r2_good}) and taking the
limit $v\to\infty$ we obtain
\begin{equation}
  \label{eq:ren_r2}
  R_2(x) = \frac14\int d\u  e^{-\pi x Q}
  e^{2i(u_1-u_2)} \left[ \Psi^2 -
    \frac{\partial \Psi}{\partial u_1} \frac{\partial
      \Psi}{\partial u_2}\right]. 
\end{equation}
The only change when the above calculation is generalized to other
boundary conditions at the central vertex (i.e.~to nonzero values of 
$\lambda^{-1}$ in (\ref{eq:SE_gen})) is the appearance of a factor 
$e^{-\lambda^{-1} (z_1 + z_2)}$ next to every occurrence of $d{\bf z}$ in the
above integrals.  For $\lambda$ fixed, this factor disappears after rescaling
$\z=\u/\beta$ and taking the limit $\beta\to\infty$.  Hence
equation (\ref{eq:ren_r2}) is then independent of $\lambda$.  In the case
when $\lambda^{-1}={\tilde \lambda}^{-1}v$, the dependence of the spectral 
statistics on the
boundary conditions at the central vertex persists.  The above expressions then
coincide with those for those for \v Seba billiards with a renormalized
coupling consant, given in \cite{BGS}.

For the derivatives of the function $\Psi$ one has 
\begin{eqnarray}
  \label{eq:Psi_der}
  \frac{\partial \Psi}{\partial u_1} &=&
  2\brr{J_0(0)\delta(u_1) +
    \sign(u_1)H(-u_1u_2)
    \frac{u_2J'_0\brr{2\sqrt{-u_1u_2}}}{\sqrt{-u_1u_2}}},\\ 
  \frac{\partial \Psi}{\partial u_2} &=&
  2\brr{-J_0(0)\delta(u_2) +
    \sign(u_1)H(-u_1u_2)
    \frac{u_1J'_0\brr{2\sqrt{-u_1u_2}}}{\sqrt{-u_1u_2}}},
\end{eqnarray}
therefore, using $J_0(0) = 1$ and $J'_0(x)=-J_1(x)$,
\begin{equation}
  \label{eq:Psi_prod_der}
  \frac{\partial \Psi}{\partial u_1} \frac{\partial \Psi}{\partial
    u_2} = -4\brr{\delta(u_1)\delta(u_2) +
    H(-u_1u_2) J_1^2\brr{2\sqrt{-u_1u_2}}}.
\end{equation}
Thus
\begin{equation}
  \label{eq:r2_almost_fin}
  R_2(x) = 1 + \!\!\int\!\! e^{-\pi xQ+2i(u_1-u_2)}
  \left[J_0^2\brr{2\sqrt{-u_1u_2}}+J_1^2\brr{2\sqrt{-u_1u_2}}\right]
  \!H(-u_1u_2) d\u. 
\end{equation}
Now we perform the change of variables $u_2 \mapsto -u_2$ arriving at
the following integral representation of the two-point correlation function,
\begin{equation}
  \label{eq:r2_fin}
  R_2(x) = 1 + \int_D e^{-\pi
    xM(\u)+2i(u_1+u_2)}
  \left[J_0^2\brr{2\sqrt{u_1u_2}}+J_1^2\brr{2\sqrt{u_1u_2}}\right]
    d\u.  
\end{equation}
Here the domain of integration $D$ includes first and third quadrants
of the $u_1u_2$-plane and $M(\u)$ is given by
\begin{eqnarray}
  \nonumber
  M(\u) &\equiv& M(u_1,u_2) = |u_1-u_2| +\int_0^{u_1}e^{i(2y-u_1+u_2)}
  \Psi(y,u_1-u_2-y)dy\\ 
  \label{M_def}
  &=& |u_1|+|u_2| - 2i\sign(u_1)\sum_{r,s=1}^\infty 
      \frac{(iu_1)^r (iu_2)^s(r+s-2)!}{r!s!(r-1)!(s-1)!}.
\end{eqnarray}

Equation (\ref{eq:r2_fin}) constitutes an exact formula for $R_2(x)$ for star 
graphs in the limit $v \rightarrow \infty$.  It is our main result.  The point
we seek to draw attention to is that it is exactly the same as the one
obtained in \cite{BGS} for \v Seba billiards 
when $|\psi_n({\bf x_0})|^2={\rm constant}$ in (\ref{eq:SE_Seba}) and 
$\lambda \rightarrow \infty$.  
We will expand on this 
observation later.   First, we consider some of the properties of the
two-point correlation function and the form factor in more detail.

%%%%%%%%%%%%%%%%%%%%%%%%%%%% New Section %%%%%%%%%%%%%%%%%%%%%%%%%%
\section{Expansion for large $x$}

To derive an expansion of the two point correlation function $R_2(x)$
for large $x$ we notice that since $M(-\u) = \bbar{M(\u)}$, the
integral over the third quadrant in (\ref{eq:r2_fin})
is equal to the complex conjugate of the integral over second
quarter-plane, i.e. 
\begin{equation}
  \label{eq:r2_x_large}
    R_2(x) = 1 + 2\Re \int\!\!\int_0^\infty  e^{-\pi
      xM(\u)+2i(u_1+u_2)} J(\u) d\u,
\end{equation}
where 
\begin{equation}
  \label{eq:expans_bessel}
  J(\u) = J_0^2(2\sqrt{u_1u_2})+J_1^2(2\sqrt{u_1u_2}) =
    \sum_{n=0}^\infty \frac{(-1)^nu_1^nu_2^n(2n)!}{(n+1)!(n!)^3}.
\end{equation}
Now we can use the expansion of $M(\u)$, (\ref{M_def}), to expand
$R_2(x)$ in the powers of $1/x$.  We substitute $u_i =
\gamma_i/(x\pi)$ and obtain
\begin{eqnarray}
  \nonumber
  R_2(x) &=& 1 + 2\Re \frac1{x^2\pi^2}\int\!\!\int_0^\infty
  d\gamma_1d\gamma_2  e^{-\gamma_1-\gamma_2} \left[1+
    \frac{2i\brr{\gamma_1+\gamma_2-\gamma_1\gamma_2}}{x\pi}\right.\\
  &\hphantom{= 1}& - \left.
    \frac{\brr{5\gamma_1\gamma_2 +
        2\gamma_1^2 + 2\gamma_2^2 - 5\gamma_1\gamma_2^2 -
        5\gamma_1^2\gamma_2 + 2\gamma_1^2\gamma_2^2}}{x^2\pi^2} +
    O\brr{\frac1{x^3}} \right] 
  \nonumber\\
  &=& 1 +  2\Re \left[\frac1{x^2\pi^2} + \frac{2i}{x^3\pi^3} -
    \frac1{x^4\pi^4} + \ldots \right].
  \label{eq:r2_exp}
\end{eqnarray}
To compare this to the expansion (\ref{eq:Ktau_firstfew}) of $K(\tau)$
we note that if $K(\tau) = 1 + \sum_{k=1}^\infty a_k \tau^k$ for
$\tau>0$ then, inverting the Fourier transform in~(\ref{eq:def_Ktau}),
\begin{eqnarray}
  \label{eq:FT_form}
  R_2(x) - 1 &=& 2 \Re \lim_{\epsilon\to0} {\int_0^\infty
    (K(\tau)-1) e^{-2\pi i(x-i\epsilon)\tau}d\tau}\\ 
  &=& 2 \Re{\sum_{k=1}^\infty
    \brr{\frac{-i}{2\pi}}^{k+1} \frac{a_k k!}{x^{k+1}} }.
\end{eqnarray}
Applying this to 
\begin{equation}
  \label{eq:K_tau_exp}
  K(\tau) = 1-4\tau+8\tau^2-\frac{8}{3}\tau^3 + O(\tau^4),
\end{equation}
we see that the first few coefficients of the two expansions
agree.  The proof that it is so for all coefficients is given by the
following proposition.
\begin{prop}	
  The asymptotic expansion (\ref{eq:r2_exp}) of the two-point correlation 
  function and the expansion~(\ref{Ktau}) of the
  form factor coincide under the Fourier transformation
  \begin{equation}
    \label{eq:agreement_theorem}
    \int\!\!\int_0^\infty  e^{-\pi xM(\u)+2i(u_1+u_2)} J(\u) d\u 
    = \int_0^\infty\brr{K(\tau')-1}e^{-2\pi ix\tau'}d\tau'.
  \end{equation}
\end{prop}

\begin{proof}
The Fourier transform in (\ref{eq:agreement_theorem})
establishes the correspondence between the terms in the asymptotic
expansion of 
\begin{equation}
  \label{eq:def_tildeR2}
  \widetilde{R_2}(x) = \int\!\!\int_0^\infty e^{-\pi
    xM(\u)+2i(u_1+u_2)} J(\u) d\u
\end{equation}
and the terms of the small $\tau$ expansion of $K(\tau)$.  This
correspondence is
\begin{equation}
  \label{eq:FT_rule}
  \frac1{(2\pi ix)^k} \longleftrightarrow
  \frac{\tau^{k-1}}{(k-1)!}.
\end{equation}

Our plan is to modify the integrand in the definition of
$\widetilde{R_2}(x)$, getting rid of the factor $e^{2i(u_1+u_2)}J(\u)$,
expand the integral in inverse powers of $x$ and apply the
correspondence rule~(\ref{eq:FT_rule}) to recover~(\ref{Ktau}). 

First of all, as one can verify by direct substitution of the series
for $M(u_1,u_2)$,
\begin{multline}
  \label{eq:dalphas_xM}
  \brr{\frac{\partial}{\partial \alpha_1} + \frac{\partial}{\partial
      \alpha_2}} \brr{xM\brr{\frac{\alpha_1}{x}, \frac{\alpha_2}{x}}}  
  =  \sum_{r,s=0}^\infty i^{r+s} \binom{r+s}{r}
  \frac{(\alpha_1/x)^r(\alpha_2/x)^s}{r!s!}\\
  =  2 e^{i(\alpha_1+\alpha_2)/x}
  J_0\brr{\frac{2\sqrt{\alpha_1\alpha_2}}{x}},
\end{multline}
and
\begin{multline}
  \label{eq:dxm_u/x}
  \frac{\partial}{\partial x} \brr{xM\brr{\frac{\alpha_1}{x},
      \frac{\alpha_2}{x}}} 
  = \sum_{r,s=1}^\infty 2i^{r+s+1}
  \frac{(r+s-1)!(\alpha_1/x)^r(\alpha_2/x)^s}{r!s!(r-1)!(s-1)!}\\
  = -\frac{2i\sqrt{\alpha_1\alpha_2}}{x}
      J_1\brr{\frac{2\sqrt{\alpha_1\alpha_2}}{x}} 
  e^{i(\alpha_1+\alpha_2)/x}.
\end{multline}
Applying (\ref{eq:dxm_u/x}),
\begin{multline}
  \label{eq:big_id1}
  \frac{\partial^2}{\partial x^2} e^{-\pi xM\brr{\frac{\alpha_1}{x},
      \frac{\alpha_2}{x}}}
  \\
  = e^{-\pi xM} \brr{-4\pi^2\frac{\alpha_1\alpha_2}{x^2}J_1^2e^{2\phi}
    - \frac{2\pi i}{x^3}\brr{2J_0e^\phi \alpha_1\alpha_2 +
      iJ_1e^\phi \sqrt{\alpha_1\alpha_2}(\alpha_1+\alpha_2)} },
\end{multline}
where $\phi=i(\alpha_1+\alpha_2)/x$ and for simplicity we have omitted the
argument \newline $(\alpha_1/x,\alpha_2/x)$ of the functions $M$,
$J_0$ and $J_1$. 

Similarly, using~(\ref{eq:dalphas_xM}), we have
\begin{multline}
  \label{eq:big_id2}
  \brr{\frac{\partial}{\partial \alpha_1} + \frac{\partial}{\partial
      \alpha_2}}^2 
  e^{-\pi xM\brr{\frac{\alpha_1}{x}, \frac{\alpha_2}{x}}}
  \\
  = e^{-\pi xM} \brr{4\pi^2J_0^2e^{2\phi}
    - \frac{2\pi i}{\alpha_1\alpha_2x}\brr{2J_0e^\phi \alpha_1\alpha_2
      + iJ_1e^\phi \sqrt{\alpha_1\alpha_2}(\alpha_1+\alpha_2)} }.
\end{multline}
Noticing the similarity between (\ref{eq:big_id1}) and
(\ref{eq:big_id2}), we subtract the first from the second, with the
appropriate factors, to obtain
\begin{multline}
  \label{eq:Bogom_id}
  \frac1{4\pi^2}
  \left[\frac{1}{x^2}\brr{\frac{\partial}{\partial\alpha_1} 
      + \frac{\partial}{\partial\alpha_2}}^2 -
    \frac{1}{\alpha_1\alpha_2} \frac{\partial^2}{\partial
      x^2}\right] 
  e^{-\pi xM\brr{\frac{\alpha_1}{x},\frac{\alpha_2}{x}}}\\ 
  = \frac{1}{x^2} \left[J_0^2+J_1^2\right] e^{2\phi} e^{-xM},
\end{multline}
where, as before, the argument $(\alpha_1/x,\alpha_2/x)$ of $M$,
$J_0$ and $J_1$ has been omitted.  The right hand side of
(\ref{eq:Bogom_id}) is exactly the 
integrand of (\ref{eq:r2_x_large}) if we perform the change of
variables $u_i=\alpha_i/x$ and, therefore,
\begin{equation}
  \label{eq:Bogom_rep}
  \widetilde{R_2}(x) 
  = \iint_0^\infty \frac{d\alpha_1 d\alpha_2}{4\pi^2} 
  \left[\frac{1}{x^2}\brr{\frac{\partial}{\partial\alpha_1} 
      + \frac{\partial}{\partial\alpha_2}}^2 -
    \frac{1}{\alpha_1\alpha_2} \frac{\partial^2}{\partial
      x^2}\right] 
  e^{-\pi xM\brr{\frac{\alpha_1}{x},\frac{\alpha_2}{x}}}.
\end{equation}

The first term in the integral can be evaluated as follows,
\begin{multline}
  \iint_0^\infty \frac{d\alpha_1 d\alpha_2}{4\pi^2x^2} 
  \brr{\frac{\partial}{\partial\alpha_1} 
    + \frac{\partial}{\partial\alpha_2}}^2 
  e^{-\pi xM\brr{\frac{\alpha_1}{x},\frac{\alpha_2}{x}}}\\
  = \left( -\int_0^\infty \frac{d\alpha_2}{4\pi x^2}
    \left[\Theta \right]_{\alpha_1=0}^\infty
    -  \int_0^\infty \frac{d\alpha_1}{2\pi x^2}
    \left[\Theta \right]_{\alpha_2=0}^\infty\right),
\end{multline}
where
\begin{equation}
  \label{eq:Theta}
  \Theta = \brr{\frac{\partial}{\partial\alpha_1} 
    + \frac{\partial}{\partial\alpha_2}} e^{-\pi
    xM\brr{\frac{\alpha_1}{x},\frac{\alpha_2}{x}}} 
  = 2 e^{i(\alpha_1+\alpha_2)/x}
  J_0\brr{\frac{2\sqrt{\alpha_1\alpha_2}}{x}} e^{-\pi xM}.
\end{equation}
Since 
\begin{equation}
  \label{eq:isaev_shtirlitz}
  \left[ \Theta \right]_{\alpha_1=0}^\infty = -2e^{i\alpha_2/x}
  e^{-\pi \alpha_2}, \qquad 
  \left[ \Theta \right]_{\alpha_2=0}^\infty = -2e^{i\alpha_1/x}
  e^{-\pi \alpha_1},
\end{equation}
we obtain
\begin{equation}
  \iint_0^\infty \frac{d\alpha_1 d\alpha_2}{4\pi^2x^2} 
  \brr{\frac{\partial}{\partial\alpha_1} 
    + \frac{\partial}{\partial\alpha_2}}^2 
  e^{-\pi xM\brr{\frac{\alpha_1}{x},\frac{\alpha_2}{x}}}
  = \frac{1}{2\pi x^2} \frac{2}{\pi-i/x}.
\end{equation}
Now we can expand the result in inverse powers of $x$ and apply
the correspondence rule (\ref{eq:FT_rule}).  We obtain
\begin{equation}
  \frac{1}{\pi x} \frac{1}{\pi x-i} 
  = -\sum_{k=0}^\infty \brr{\frac{i}{\pi x}}^{k+2}
  \longleftrightarrow 2 \sum_{k=0}^\infty
  \frac{(-2\tau)^{k+1}}{(k-1)!}
  = 2\brr{e^{-2\tau}-1}.
\end{equation}

Next we need to expand the second part of the integrand in
(\ref{eq:Bogom_rep}),
\begin{multline}
  \label{eq:second_part}
  \frac{\partial^2}{\partial x^2} 
  e^{-\pi xM} = 
  \frac{\partial^2}{\partial x^2} e^{-\pi(\alpha_1+\alpha_2)}
  \exp\brr{2\pi i\sum_{r,s=0}^\infty 
    \frac{(i\alpha_1)^{r+1} (i\alpha_2)^{s+1}(r+s)!}
    {x^{r+s+1}r!s!(r+1)!(s+1)!} }\\ 
  = e^{-\pi(\alpha_1+\alpha_2)}\frac{\partial^2}{\partial x^2} 
  \left[ \sum_{j=0}^\infty \frac{(2\pi i)^j}{j!} 
    \brr{\sum_{r,s=0}^\infty 
      \frac{(i\alpha_1)^{r+1} (i\alpha_2)^{s+1}(r+s)!}
      {x^{r+s+1}r!s!(r+1)!(s+1)!}}^j
  \right].
\end{multline}
Using the same notation as in (\ref{eq:notation_F}),
\begin{multline}
  \brr{\sum_{r,s=0}^\infty
    \frac{(i\alpha_1)^{r+1}
    (i\alpha_2)^{s+1}(r+s)!}{x^{r+s+1}r!s!(r+1)!(s+1)!}}^j 
  = \brr{\sum_{r,s=0}^\infty
    \frac{(i\alpha_1)^{r+1} (i\alpha_2)^{s+1}}{x^{r+s+1}}F_1(r,s)}^j\\
  = \sum_{R,S=0}^\infty \frac{(i\alpha_1)^{R+j}
    (i\alpha_2)^{S+j}}{x^{R+S+j}}  F_j(R,S),
\end{multline}
where, as before, $F_j(R,S)$ is the $j$th convolution of $F_1(R,S)$
with itself.  Thus
\begin{multline}
  \label{eq:second_part_cont}
  \frac{\partial^2}{\partial x^2} 
  e^{-\pi xM\brr{\frac{\alpha_1}{x},\frac{\alpha_2}{x}}} 
  =  e^{-\pi(\alpha_1+\alpha_2)} \sum_{j=1}^\infty 
  \frac{(2\pi i)^j}{j!}\\ 
  \times \sum_{R,S=0}^\infty \frac{(R+S+j-1)!(i\alpha_1)^{R+j} 
    (i\alpha_2)^{S+j}}{(R+S+j+1)!x^{R+S+j+2}} F_j(R,S). 
\end{multline}
Finally we integrate against
$d\alpha_1d\alpha_2/(4\pi^2\alpha_1\alpha_2)$ to arrive at
\begin{multline}
  - \iint_0^\infty \frac{d\alpha_1
    d\alpha_2}{4\pi^2\alpha_1\alpha_2}  
  \frac{\partial^2}{\partial x^2} 
  e^{-\pi xM\brr{\frac{\alpha_1}{x},\frac{\alpha_2}{x}}}\\
  =  -\sum_{j=1}^\infty \frac{(2\pi i)^j}{4\pi^2j!} 
  \sum_{R,S=0}^\infty \frac{(R+S+j+1)!(R+j-1)!(S+j-1)!}
  {(R+S+j-1)!(-i\pi)^{R+S+2j}x^{R+S+j+2}} F_j(R,S) \\
%  = \sum_{j=1}^\infty \frac{(-2)^j}{4j!} 
%  \sum_{R,S=0}^\infty \frac{(R+S+j+1)!(R+j-1)!(S+j-1)!}
%  {(R+S+j-1)!(-i\pi x)^{R+S+j+2}} F_j(R,S)\\
  \longleftrightarrow 
%  \sum_{j=1}^\infty \frac{(-2)^j}{4j!} 
%  \sum_{R,S=0}^\infty \frac{(-2\tau)^{R+S+j+2}(R+j-1)!(S+j-1)!}
%  {\tau(R+S+j-1)!} F_j(R,S)\\
  \tau \sum_{j=1}^\infty \frac{(4\tau)^j}{j!} 
  \sum_{R,S=0}^\infty \frac{(-2\tau)^{R+S}(R+j-1)!(S+j-1)!}
  {(R+S+j-1)!} F_j(R,S).
\end{multline}
This is exactly the same as the $j$ sum in (\ref{Ktau}) with the
exception of the extra $j=1$ term in the summation above.  For $j=1$
we have 
\begin{multline}
  4\tau^2 \sum_{R,S=0}^\infty \frac{(-2\tau)^{R+S}R!S!}
  {(R+S)!} F_j(R,S) 
  = \sum_{R,S=0}^\infty \frac{(-2\tau)^{R+S+2}}{(R+1)!(S+1)!}\\
  = \brr{\sum_{R=0}^\infty \frac{(-2\tau)^{R+1}}{(R+1)!}}
  \brr{\sum_{S=0}^\infty \frac{(-2\tau)^{S+1}}{(S+1)!}} =
  (1-e^{-2\tau})^2\\
  = 1-2e^{-2\tau} + e^{-4\tau},
\end{multline}
which, together with the terms $1$ and $2(e^{-2\tau}-1)$,
gives the correct contribution $e^{-4\tau}$.
\end{proof}

%%%%%%%%%%%%%%%%%%%%% New Section %%%%%%%%%%%%%%%%%%%%%%%%%%%
\section{Singularities of the form factor}

One can also obtain some information about the
singularities of $K(\tau)$ by Fourier transforming the integral representation
(\ref{eq:r2_x_large}).   There is, however, a subtle
problem associated with this approach.  The form factor is by definition an
even function defined on the real line.  What we want to get from
transforming (\ref{eq:r2_x_large}) is an analytic function which
coincides with the form factor for real $\tau>0$, so as to be able to
study its complex singularities. 

As we saw above,
\begin{equation}
  \label{eq:R2tilde}
  \widetilde{R_2}(x) = \int\!\!\int_0^\infty e^{-\pi
    xM(\u)+2i(u_1+u_2)} J(\u) d\u = \int_0^\infty (K(\tau')-1)
    e^{-2\pi ix\tau'}. 
\end{equation}
Integrating (\ref{eq:R2tilde}) against $e^{2\pi ix\tau}$ on the real
line we obtain
\begin{equation}
  \label{eq:invR2tilde}
  \int_{-\infty}^\infty \widetilde{R_2}(x) e^{2\pi ix\tau} dx =
  K(\tau) - 1, \qquad \tau>0.
\end{equation}
One can check that this leads to the correct power series expansion
of the form factor: give $x$ a small negative imaginary part,
$x\mapsto x-i\epsilon$, in $\widetilde{R_2}(x)$ (this is
consistent with (\ref{eq:R2tilde})), substitute in the asymptotic
expansion (56), and integrate term-by-term.   

We now use $\widetilde{R_2}(-x)=\bbar{\widetilde{R_2}(x)}$ to
write 
\begin{equation}
  \int_{-\infty}^\infty e^{2\pi ix\tau} \widetilde{R_2}(x) dx 
  = \int_0^\infty  \brr{e^{2\pi ix\tau}\widetilde{R_2}(x) + e^{-2\pi
  ix\tau}\bbar{\widetilde{R_2}(x)}} dx. 
  \label{eq:pFT}
\end{equation}
The only factor $\widetilde{R_2}(x)$ which depends on $x$ is $e^{-\pi
  xM(\u)}$ and
\begin{equation}
  \label{eq:FT_exM}
  \int_0^\infty e^{2\pi ix\tau} e^{-\pi xM(\u)} dx =
  \frac{1}{\pi(M(\u)-2i\tau)}, 
\end{equation}
thus we have for the form factor
\begin{equation}
  \label{eq:int_rep_ff}
  K(\tau) = 1 + \frac1\pi \int\!\!\int_0^\infty \left[
  \frac{e^{2i(u_1+u_2)}}{M(\u)-2i\tau} + 
  \frac{e^{-2i(u_1+u_2)}}{\bbar{M(\u)}+2i\tau}\right] J(\u) d\u.
\end{equation}

The representation~(\ref{eq:int_rep_ff}) presents us with a way to
find the singularities of $K(\tau)$.  These are given
by the condition $\tau=M(\u_s)/(2i)$ and $\tau=\bbar{M(\u_s)/(2i)}$,
where 
the point $\u_s$ is such that 
\begin{equation}
  \label{eq:sing_cond}
  \frac{\partial M}{\partial u_1}(\u_s) = 
  \frac{\partial M}{\partial u_2}(\u_s) = 0.
\end{equation}
The derivative with respect to $u_2$ is
\begin{equation}
  \label{eq:deriv_u2}
  \frac{\partial M}{\partial u_2} = 1 - 2\int_0^{u_1} \left[
   e^{i(y+z)} J_1\brr{2\sqrt{yz}} \sqrt{y/z} - i e^{i(y+z)}
  J_0\brr{2\sqrt{yz}} \right] dy,
\end{equation}
where $z=y-u_1+u_2$ and we have assumed that $u_1>u_2>0$.  It is obvious
from the expansion~(\ref{M_def}), however, that the function $M(\u)$
is continuously differentiable if 
$u_1u_2>0$ and hence that the expression~(\ref{eq:deriv_u2}) is valid for
all $u_1>0$ and $u_2>0$.
The integral in (\ref{eq:deriv_u2}) is not easy to analyse and
to simplify it we reduce our search to the line $u_2=u_1$, where
\begin{equation}
  \frac{\partial M}{\partial u_2}(u_2=u_1) = 1 -
  2\int_0^{u_1}e^{2iy}J_1(2y)dy + 2i\int_0^{u_1} e^{2iy}J_0(2y)dy.
\end{equation}
Performing the second integration by parts,
\begin{equation}
  \label{eq:int_parts}
  \int_0^{u_1} e^{2iy}J_0(2y)dy =
  \left.\frac{e^{2iy}J_0(2y)}{2i}\right|_0^{u_1} +
  \frac{2}{2i}\int_0^{u_1}e^{2iy}J_1(2y)dy,
\end{equation}
we obtain, after simplification,
\begin{equation}
  \label{eq:deriv_res}
  \frac{\partial M}{\partial u_2}(u_2=u_1) = e^{2iu_1}J_0(2u_1).
\end{equation}
Since
$ \frac{\partial M}{\partial u_1}(u_2=u_1) = \frac{\partial
  M}{\partial u_2}(u_2=u_1)$, we see that 
the zeros of the derivatives of $M(\u)$
on the line $u_2=u_1$ are given by the zeros of the Bessel function
$J_0$.  The nearest zero is at $u_s\approx 1.202$.  
Thus one of the singularities of $K(\tau)$ lies at
$\tau_s = M(1.202,1.202)/(2i) = 0.462-0.420 i$.  We note that
$|\tau_s|=0.624$, which coincides with our previous numerical estimate
of the radius of convergence of
the series expansion of $K(\tau)$ in powers of $\tau$ around $\tau=0$.  
This strongly suggests that this singularity is the closest to the
origin.  To this end, we can prove the following.

\begin{prop}
  \label{prop:nearest_sing}
  Among the singularities arising from stationary points of\linebreak
  $M(u_1, u_2)$ along the line $u_2=u_1$, 
  the singularity at $\tau_s = M(1.202,1.202)/(2i)$ is the
  nearest to the origin.
\end{prop}
\begin{proof}
  To show that the statement is true we need to prove that the
  function $|M(u,u)|^2$ is a nowhere decreasing function of
  $u$.  On the line
  $u_1=u_2=u$ we have
  \begin{equation}
    \label{eq:M_on_line}
    M(u,u) = \int_0^{2u} e^{iy}J_0(y) dy =
    2e^{2iu}u\brr{J_0(2u)-iJ_1(2u)}. 
  \end{equation}
  Thus $|M(x/2,x/2)|^2 = x^2\brr{J_0^2(x)+J_1^2(x)}$ and its
  derivative is, after simplification, $\frac{d}{dx}|M(x/2,x/2)|^2 =
  2xJ_0^2(x) \geq 0$.
\end{proof}

\begin{figure}[t]
\vskip 8.5cm
\includegraphics{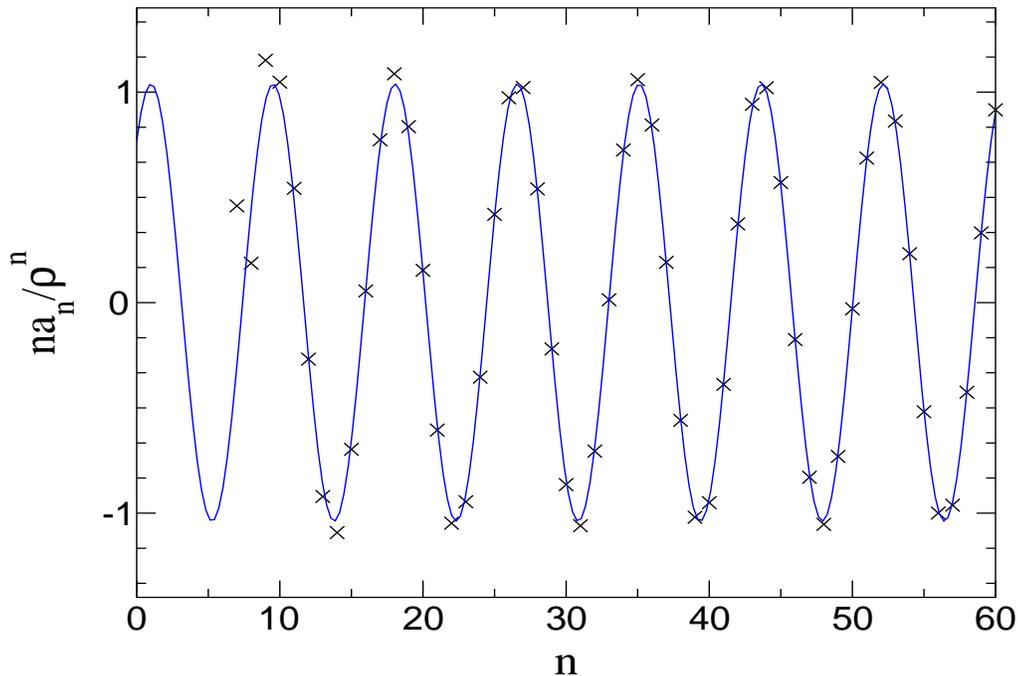}
\caption{The coefficients of the power series expansion of $K(\tau)$
  normalized by $\rho^n$ (crosses), compared to 
  (\ref{eq:leading_contrib_ser_K}).  As expected, the agreement 
  improves as $n$
  increases.} 
\label{fig:cosfit}
\end{figure}

It is straightforward to approximate the behaviour of $K(\tau)$ near
these singularities.  We expand 
\begin{eqnarray}
  \nonumber
  M(\u) &\approx& M(\u_s) + \frac12 \frac{\partial^2 M}{\partial
    u_1^2}(\u_s) (u_1-u_2)^2 + \frac12 \frac{\partial^2 M}{\partial
    u_2^2}(\u_s) (u_2-u_s)^2\\
  \nonumber
  &&\qquad + \frac{\partial^2 M}{\partial u_1\partial u_2}(\u_s)
  (u_1-u_s)(u_2-u_s)\\ 
  &=& M(\u_s) + \alpha_s\brr{(u_1-u_s)^2+(u_2-u_s)^2}.
  \label{eq:M_expand_sing}
\end{eqnarray}
For the singularity associated with the first Bessel zero, 
$\alpha_s\approx 0.385-0.349 i$.  Then, when $\tau$ is real, 
\begin{equation}
  \label{eq:ff_around_sing}
  K(\tau)\! \approx \!\frac1{\pi\alpha_s} \iint_{0}^\infty
  \frac{J(\u) e^{2i(u_1+u_2)}d\u}{(u_1-u_s)^2 + (u_2-u_s)^2 +
    (M(\u_s)-2i\tau)/\alpha_s} + \mbox{c.c.}
\end{equation}
The main
contribution to the integral around these singularities is
\begin{equation}
  \label{eq:ln_sing}
  K(\tau) \propto -C\ln\brr{1-\frac{2i\tau}{M(\u_s)}} -
  \bbar{C}\ln\brr{1+\frac{2i\tau}{\bbar{M(\u_s)}}}, 
\end{equation}
where $C=J(\u_s)e^{4iu_s}/\alpha_s$.  Expanding
(\ref{eq:ln_sing}) into a series around $\tau=0$ we get 
\begin{equation}
  \label{eq:leading_contrib_ser_K}
  K(\tau) \propto 2\Re\brr{C\sum_{n=1}^\infty
    \rho^n\frac{e^in\phi}{n}\tau^n} = 2A\sum_{n=1}^\infty\cos(\phi n + \psi)
  \frac{\rho^n}{n} \tau^n,
\end{equation}
where, for the singularity analysed above, 
$A = \left|J(\u_s)e^{4iu_s}/\alpha_s\right| \approx 0.519$,
$\psi = \arg\brr{J(\u_s)e^{4iu_s}/\alpha_s}\approx -0.737$, $\rho =
|2i/M(\u_s)|\approx 1.602$ and $\phi = \arg\brr{2i/M(\u_s)}\approx
0.737$.  By Darboux's Principle, the coefficients of the 
expansion (\ref{eq:leading_contrib_ser_K}) should
comprise the leading contribution to large-order asymptotics of 
the exact coefficients given by
(\ref{Ktau}) and (\ref{eq:CM1}).  To compare them we plot the
exact coefficients $na_n/\rho^n$ against the approximate coefficients
$2A\cos(\phi n + \psi)$.  The result is shown in Fig.~\ref{fig:cosfit}.

%%%%%%%%%%%%%%%%%%%%%%%%%%% New Section %%%%%%%%%%%%%%%%%%%%%%%%%

\section{Small $x$ limit of $R_2(x)$}
\label{sec:int_smallx}

Returning to (\ref{eq:ren_r2}), one can check that 
the function $\Psi$, defined by (\ref{eq:Psi}),  satisfies the
equation  
\begin{multline}
  \label{eq:PDPhi}
  \left[\frac{\partial^2}{2\partial u_1 \partial u_2} +
    i\brr{\frac{\partial}{\partial u_1} -
      \frac{\partial}{\partial u_2}}\right]
  \brr{e^{2i(u_1-u_2)}\Psi^2}\\
  = e^{2i(u_1-u_2)} \brr{\frac{\partial\Psi}{\partial z_1}
    \frac{\partial\Psi}{\partial z_2} - \Psi^2}.
\end{multline}
Substituting it into (\ref{eq:ren_r2}) and integrating by parts we
obtain 
\begin{multline}
  \label{eq:r2_better}
  R_2(x) = - \frac14\int d\u  e^{-\pi x Q}
  \left[\frac{\partial^2}{2\partial u_1 \partial u_2} +
    i\brr{\frac{\partial}{\partial u_1} -
      \frac{\partial}{\partial u_2}}\right]
  \brr{e^{2i(u_1-u_2)}\Psi^2} \\
  = \int \frac{d\u}{4} e^{2i(u_1-u_2)}\Psi^2
  \left[i\brr{\frac{\partial}{\partial u_1} -
      \frac{\partial}{\partial u_2}}-\frac{\partial^2}{2\partial
    u_1 \partial u_2}\right] \brr{e^{-\pi x Q}}.
\end{multline}
Now, using the identities 
\begin{equation}
  \label{eq:iden_Q}
  \frac{\partial Q}{\partial u_1} - \frac{\partial Q}{\partial u_2} = 
  e^{i(u_1-u_2)}\Psi, \qquad 
  \frac{\partial^2 Q}{2\partial u_1 \partial u_2} = 
  -i e^{i(u_1-u_2)}\Psi,
\end{equation}
which one can derive using the series expansion of
$Q(u_1,u_2)=M(u_1,-u_2)$, we write 
\begin{multline}
  \label{eq:diff_op_eQ}
  \left[i\brr{\frac{\partial}{\partial u_1} -
      \frac{\partial}{\partial u_2}}-\frac{\partial^2}{2\partial
      u_1 \partial u_2}\right] \brr{e^{-\pi x Q}}\\ 
  = e^{-\pi x Q} \brr{ -i\pi x\brr{\frac{\partial Q}{\partial u_1} -
      \frac{\partial Q}{\partial u_2}} + \frac{\pi x}2 \frac{\partial^2
      Q}{\partial u_1 \partial u_2} - 
    \frac{(\pi x)^2}2 \frac{\partial Q}{\partial u_1} 
    \frac{\partial Q}{\partial u_2}}\\ 
  = -e^{-\pi x Q} \brr{\frac{3i\pi x}{2}e^{i(u_1-u_2)}\Psi +
    \frac{(\pi x)^2}2 \frac{\partial Q}{\partial u_1} 
    \frac{\partial Q}{\partial u_2} }.
\end{multline}
Thus we obtain, finally,
\begin{equation}
  \label{eq:r2_smallx}
  R_2(x) = -\int \frac{d\u}{8}
  e^{2i(u_1-u_2)-\pi x Q}\Psi^2 \left[ \pi^2x^2\frac{\partial
  Q}{\partial u_1}\frac{\partial Q}{\partial u_2} +
   3i\pi x \Psi e^{i(u_1-u_2)}\right].
\end{equation}
From (\ref{eq:r2_smallx}) one can see that the two-point
correlation function $R_2(x)$ is linear in $x$ for small $x$. The
slope was computed in \cite{BGS}:
\begin{equation}
  \label{eq:R_2_small_x}
  R_2(x) = \frac{\pi\sqrt{3}}{2}x + O(x^2).
\end{equation}

%%%%%%%%%%%%%%%%%%%%%%%%%%%%%%%%%%%%%%%%%%%%%%%%%%%%%%%%%%%%%%%%%%%%
\section{Discussion}

The derivation presented above provides a proof that two-point spectral
correlations for certain \v Seba billiards and quantum star graphs
are the same, in the appropriate limits.  This
initially surprising fact has its explanation in the following 
observations.  First, the dynamics in both systems is centered around a
single point scatterer; in star graphs it is the central vertex,
and in \v Seba billiards the singularity.  Furthermore, in between 
scatterings the dynamics is integrable in both cases.

Second, applying the Mittag-Leffler
theorem to the meromorphic function $\tan z$, we have that
\begin{equation}
  \label{eq:mitt_leff}
  \tan z = \sum_{n=-\infty}^\infty \brr{\frac1{n\pi+\pi/2-z} -
  \frac1{n\pi+\pi/2}}.
\end{equation}
We can therefore rewrite (\ref{eq:SE_gen}) in a form similar to
(\ref{eq:Sebacond}) when $|\psi_n({\bf x_0})|^2={\rm constant}$.  It 
thus becomes less surprising that the 
two point correlation functions of the two systems are the same, 
because in the limit
$v\to\infty$ the poles in (\ref{eq:SE}) have properties 
similar to those of a Poisson sequence.  
%? sequence'

Third, from the mathematical point of view star graphs and \v Seba billiards
are similar in that in both cases the scattering centre corresponds
quantum mechanically to a perturbation of rank one.

Finally, we remark that our results demonstrate that, at least 
as regards the
special case considered here,  graphs are able to reproduce 
features of other, experimentally realizable, quantum systems, and also that
they provide further confirmation that spectral statistics can be computed
exactly using the trace formula when the periodic orbit statistics are
known \cite{BK}.

\section*{Acknowledgments}
One of us (G.B.) would like to thank the 
Laboratoire de Physique Th\'eorique et Mod\`eles Statistiques, 
Universit\'e Paris-Sud,
and BRIMS, Hewlett-Packard Laboratories Bristol,
for their hospitality.

\end{document}